\documentclass[aps,floatfix,nofootinbib,preprintnumbers,amsmath, amssymb,showkeys,preprint]{revtex4-1}

\usepackage{epsf,epsfig,subfigure,graphicx,amsmath,amssymb,epstopdf,color}

\begin{document}

\preprint{KIAS-P15030}
\preprint{CETUP2015-008}

\title{\bf Electro-Weak Dark Matter:\\
non-perturbative effect confronting indirect detections
}

\author{
Eung Jin Chun$^{(1)}$\footnote{email: ejchun@kias.re.kr} and Jong-Chul Park$^{(2)}$\footnote{email: log1079@gmail.com}
}
\affiliation{
$^{(1)}$ Korea Institute for Advanced Study, Seoul 130-722, Korea \\
$^{(2)}$ Department of Physics, Sungkyunkwan University, Suwon 440-746, Korea
}

\begin{abstract}
We update indirect constraints on Electro-Weak Dark Matter (EWDM) considering the
Sommerfeld-Ramsauer-Townsend (SRT) effect for its annihilations into a pair of standard model gauge bosons assuming that EWDM accounts for the observed dark matter (DM) relic density for a given DM mass and mass gaps among the multiplet components. For the radiative or smaller mass splitting, the hypercharged triplet and higher multiplet EWDMs are ruled out up to the DM mass $\approx$ 10 -- 20 TeV by the combination of the most recent data from AMS-02 (antiproton),
Fermi-LAT (gamma-ray), and HESS (gamma-line).
The Majorana triplet (wino-like) EWDM can evade all the indirect constraints
only around Ramsauer-Townsend dips which can occur for a tiny mass splitting of order 10 MeV or less.
In the case of  the doublet (Higgsino-like) EWDM, a wide range of its mass $\gtrsim$ 500 GeV is allowed except Sommerfeld peak regions.
Such a stringent limit on the triplet DM can be evaded by employing a larger mass gap of the order of 10 GeV which allows its mass larger than about 1 TeV.
However, the future CTA experiment will be able to cover most of the unconstrained parameter space.
\end{abstract}

\maketitle

\section{Introduction}\label{Intro}

The most simplistic candidate for DM would be a neutral component of an ${\rm SU(2)}_L \times {\rm U(1)}_Y$ multiplet  added to the Standard Model (SM) \cite{Hisano:2003ec,Cirelli:2005uq}, which is dubbed as EWDM.
In the case of a fermionic EWDM, its physical properties are completely determined by the gauge charge, dark matter mass, and mass differences between the multiplet components.
As is well-known, the non-perturbative effect plays a crucial role when the DM is slowly moving and the DM
mass is larger than the force-carrying boson mass \cite{Hisano:2003ec, Hisano:2004ds, Cirelli:2007xd, ArkaniHamed:2008qn,Chun:2008by, Chun:2012yt}.
Furthermore, this effect includes not only the usual Sommerfeld enhancement but also the Ramsauer-Townsend suppression which are more apparent for larger DM mass or smaller mass gaps \cite{Chun:2012yt}.

In this paper, we revisit the non-perturbative effect of EWDM which is strongly constrained by the recent indirect detection data on anti-proton flux by AMS-02 \cite{AMS2015}, gamma-ray measurement from Milky Way (MW) satellite dwarf galaxies by Fermi Large Area Telescope (Fermi-LAT)~\cite{Ackermann:2015zua}, and gamma-ray line searches by Fermi-LAT \cite{TheFermi-LAT:2015gja} and High Energy Spectroscopic System (HESS)~\cite{Abramowski:2013ax}.
The first two measurements put bounds on the leading-order annihilation processes: $\chi^0 \chi^0 \to WW/ZZ$, and the last two on the loop-induced ones: $\chi^0 \chi^0 \to \gamma \gamma/ \gamma Z$.
Remark that the updated Fermi-LAT search on gamma-ray lines covers the dark matter mass up to around 500 GeV complimenting the previous HESS search range of 500 GeV -- 25 TeV.
Related studies have been made for the case of the wino or wino-Higgsino dark matter in a supersymmetric theory \cite{Hryczuk:2010zi, Cohen:2013ama, Fan:2013faa, Hryczuk:2014hpa,  Beneke:2014hja, Ibe:2015tma}.

The annihilation of two neutral particles into $\gamma +X$  is a radiative process and may be subject to a large correction due to the resummation of electroweak Sudakov logarithms $\alpha_2 \log^2(m_{\rm DM}^2/m_W^2)$.
In Refs.~\cite{Bauer:2014ula, Ovanesyan:2014fwa}, the authors studied the Sudakov resummation effect for the annihilation rate of the exclusive process $\chi^0 \chi^0 \to \gamma \gamma/\gamma Z$ and found that the exclusive rate is reduced by up to a factor of 2 -- 3 compared to the tree-level plus Sommerfeld enhancement calculation.
On the other hand, Refs.~\cite{Baumgart:2014vma} treated systematically the semi-inclusive annihilation rate into the final state $\gamma + X$ within the resolution of the experiment to find that the effect of higher order correction is very limited as the semi-inclusive rate changes by only $\mathcal{O}(1 \%)$ at $m_{\rm DM}=3$ TeV.
Thus, we will simply use the leading-order annihilation cross section of EWDM including only the SRT effect.

For our study, we will consider three specific examples of fermionic EWDM in the lowest-lying multiplets:
a vector-like (Dirac) doublet with a hypercharge $Y=\pm 1/2$
(Higgsino-like), a (Majorana) triplet with $Y=0$ (wino-like), and a
vector-like (Dirac) triplet with $Y=\pm1$.
Note that a certain symmetry like $Z_2$ has to be imposed for the stability of these EWDM candidates.
They are assumed to form 100 \% of the observed DM density in wide ranges of the DM mass and mass gaps among the multiplet components.
As shown in Ref.~\cite{Chun:2012yt}, the feature of Sommerfeld peaks and Ramsauer-Townsend dips in the non-perturbative annihilation cross section depends sensitively on those mass parameters, which leads to an interesting impact on the indirect detection constraints.

This paper is organized as follows.
In Section II, we review the non-perturbative effect, the SRT effect for annihilations of EWDM summarizing the results in Ref.~\cite{Chun:2012yt}.
In Section III, we discuss the existing indirect constrains on annihilations of EWDM into SM gauge bosons based on various cosmic-ray measurements, then conclude in Section IV.
Finally, in Appendix A, we present the explicit forms of the potential and scattering matrices described in Section II.

\section{EWDM and Sommerfeld-Ramsauer-Townsend effect}

Let us first make a brief summary of the non-perturbative effect on the EWDM annihilation~\cite{Chun:2012yt}.
The doublet EWDM (Higgsino-like) consists of a Dirac fermion pair denoted by $\chi_u = (\chi_u^+, \chi_u^0)$ and $\chi_d= (\chi^0_d, \chi^-_d)$ in the chiral
representation.
The lighter linear combination of $\chi_u^0$ and $\chi_d^0$ is the DM candidate and
the mass splitting can come from (effective) non-renormalizable operators after the electroweak symmetry breaking.
The wino-like EWDM multiplet, a triplet with $Y=0$ is denoted by $\chi=(\chi^+, \chi^0, \chi^-)$ having only one Majorana neutral component.
Finally,
the triplet EWDM multiplet with $Y=\pm1$ consists of two chiral fermions, $\chi_u=(\chi_u^{++}, \chi_u^+, \chi_u^0)$ and $\chi_d=(\chi_d^{0}, \chi_d^{-},
\chi_d^{--})$.
Recall that the electroweak one-loop correction generates a mass splitting of order 0.1 GeV between the multiplet components \cite{Cirelli:2005uq}.
Together with the above-mentioned (tree-level) contribution, this one-loop correction can make arbitrary mass gaps as assumed in our analysis.

The non-perturbative effect in the EWDM annihilation arises from the exchange of the electroweak gauge bosons which mixes together the two-body states of the multiplet
components.
In the case of the doublet EWDM, there are three states formed by the charged (Dirac) component and two neutral (Majorana) components:
$\chi_u^+ \chi_d^-$, $\chi_1^0 \chi_1^0$, and $\chi^0 \chi^0$, where $\chi^0$ denotes the dark matter component.
For the wino-like EWDM, there are two two-body states:
$\chi^+ \chi^-$ and $\chi^0\chi^0$.
The triplet EWDM with $Y=\pm1$ has four two-body states:
$\chi_u^{++}\chi_d^{--}$, $\chi_u^+ \chi_d^-$, $\chi_1^0 \chi_1^0$, and $\chi^0 \chi^0$.

For the calculation of the non-perturbative effect, we apply the analysis of Refs.~\cite{Hisano:2003ec, Hisano:2004ds, Cirelli:2007xd} in which
the Green's functions $g_{ij}$ corresponding to the transition between the two-body states $i$ and $j$ are shown to follow the Schr\"odinger equation:
\begin{equation} \label{schroedinger}
 -{1\over m_{\rm DM}} {\partial^2 g_{ij} (r) \over \partial r^2} + V_{ik}(r) g_{kj} (r) = K g_{ij}(r)\,,
\end{equation}
with the boundary condition $g_{ij}(0)=\delta_{ij}$ and $\partial
g_{ij}(\infty) /\partial r = i \sqrt{m_{\rm DM}
(K-V_{ii}(\infty))}g_{ij}(\infty)$.
Here $K=m_{\rm DM} \beta^2$ is the total kinetic energy of the two initial dark matter particles in the annihilation process, where $\beta$ is the DM velocity.
Then, the non-perturbative annihilation cross section of the dark matter $\chi^0$ is
\begin{equation} \label{sigmaDM}
 \sigma v ( \chi^0 \chi^0 \to A B) = 2 d_{0i} d^*_{0j} \Gamma_{ij}^{AB}\,,
\end{equation}
where $d_{0j} = g_{0j}(\infty)$ and $v=2 \beta$ is the relative velocity between the two incident DM particles.
Here $A$ and $B$ run over the gauge bosons $(W^+, W^-, Z, \gamma)$, and the gauge
boson final states $AB$ can be $W^+ W^-$, $ZZ$,  $\gamma Z$, or  $\gamma\gamma$.
The explicit forms of the potential matrix $V_{ij}$ in Eq.~(\ref{schroedinger}) and the scattering matrix $\Gamma_{ij}^{AB}$ in Eq.~(\ref{sigmaDM}) are collected in Appendix \ref{appendix}.

The non-perturbative calculation of the EWDM annihilation exhibits not only the usual Sommerfeld enhancement with resonance peaks,
but also a vanishing cross section realizing the Ramsauer-Townsend effect for particular choices of the model parameters.
These effects are particularly sensitive to the the mass splittings between the dark matter and the charged components of the multiplet, and can appear even for the DM mass below the TeV scale when the mass splitting is reduced to $\mathcal{O}$(10) MeV or less.
One can find a detailed analysis on non-perturbative effects on the annihilation cross section of the EWDM in Ref.~\cite{Chun:2012yt}.
The appearance of the Ramsauer-Townsend dips is of a particular interest as it can allow the EWDM to evade strong constraints from various indirect detections.

\section{Constraints on EWDM from indirect DM searches}

\subsection{Annihilation into $WW$ $\&$ $ZZ$ final states}

We first present constraints on EWDM from indirect signals by its annihilation into $WW/ZZ$ final states.
Annihilations of EWDM are expected to produce $W/Z$ bosons plentifully since EWDM is charged under the SU(2) symmetry of the SM.
Fragmentation of the produced $W/Z$ bosons leads to sizable contributions to the antiproton flux and the continuum photon spectrum which are detectable in cosmic-ray measurement experiments such as AMS-02 and Fermi-LAT.

\subsubsection{Constraints from AMS-02 antiproton flux measurements}

Antiproton production from DM annihilations into $WW$ and $ZZ$ channels is constrained by the precise measurements on antiproton flux by AMS-02~\cite{AMS2015}.
The antiproton contribution from DM annihilation should be summed to the secondary antiprotons, produced by collisions of energetic cosmic-rays with the interstellar medium, which account for the bulk of the observed flux.
Ref.~\cite{Giesen:2015ufa} provided constraints on DM annihilation cross sections for various final states including $WW$ based on the recent antiproton to proton ratio measurements by AMS-02~\cite{AMS2015}.
The authors of Ref.~\cite{Giesen:2015ufa} assumed the Einasto DM halo profile and the MED propagation parameter set proposed in Ref.~\cite{Donato:2003xg}.
They also showed that for $m_{\rm DM} > 100$ GeV the limits are $\sim 2$ times weaker for the Burkert profile and $2-3$ times more stringent for the MAX propagation model.\footnote{In a number of recent papers based on Galactic synchrotron emission~\cite{Bringmann:2011py, DiBernardo:2012zu, Orlando:2013ysa, Fornengo:2014mna} and cosmic-ray positrons~\cite{DiBernardo:2012zu, DiMauro:2014iia, Lavalle:2014kca}, it has been pointed out that the thin halo model in the MIN propagation scheme is seriously disfavored.}
In our study, we use the $2\sigma$ exclusion bound on the $WW$ channel for the Einasto profile with the MED propagation model in Ref.~\cite{Giesen:2015ufa} as the bound on the total cross section $\sigma v^{WW} + \sigma v^{ZZ} \equiv \sigma v_{\chi^0\chi^0 \to W^+W^-} + \sigma v_{\chi^0\chi^0 \to ZZ}$ since the antiproton yields per annihilation from two final states $WW$ and $ZZ$ are almost undistinguishable.
The AMS-02 antiproton flux limit on $\sigma v^{WW} + \sigma v^{ZZ}$ is shown as a red-thin dotted line in Figures \ref{Fig-Higgsino}--\ref{Fig-Triplet1-BigGap}.

\subsubsection{Fermi-LAT continuum photon constraints: Milky Way satellite dwarf galaxies}

The dwarf spheroidal satellite galaxies of the Milky Way are some of the most promising targets for the DM indirect detection via gamma-rays since they are highly DM dominated objects with a relatively short distance from the Earth.
In EWDM annihilation, the continuum photons originate mostly from fragmentation of hadronic final states in the $\chi^0\chi^0 \to WW/ZZ$ processes which are strongly constrained by the gamma-ray measurements from the MW satellite dwarf galaxies~\cite{Ackermann:2015zua}.
The total mass of a dwarf galaxy within the half-light radius
and the integrated J-factor, the line-of-sight integral through the DM distribution integrated over the solid angle for a region of interest, have been found to be pretty insensitive to the change of a DM density profile~\cite{Martinez:2009jh, Strigari:2013iaa, Ackermann:2013yva}.
Thus in Ref.~\cite{Ackermann:2015zua}, the bound from satellite dwarf galaxies is obtained based on 6 years of Fermi-LAT data processed with the Pass 8 event-level analysis just assuming that the DM distribution in dwarf galaxies follows a Navarro-Frenk-White (NFW) profile~\cite{Navarro:1996gj}.
In our analysis, we take the Fermi-LAT continuum photon limit on $\sigma v^{WW} (+ \sigma v^{ZZ})$ at $2\sigma$ level for the MW satellite dwarf galaxies from Ref.~\cite{Ackermann:2015zua} which is shown as a red-thin dashed line in Figures~\ref{Fig-Higgsino}--\ref{Fig-Triplet1-BigGap}.

\subsection{Annihilation into $\gamma\gamma$ $\&$ $\gamma Z$ final states}

In this subsection, we provide indirect constraints on EWDM by its annihilation into $\gamma\gamma/\gamma Z$ final states.
Monochromatic photons arise from the one-loop processes $\chi^0 \chi^0 \to \gamma\gamma/\gamma Z$.
Such a line signature would be quite easily distinguished from astrophysical
photon sources since in most cases they produce continuous spectra.
Now, the annihilation cross sections for the processes $\chi^0 \chi^0 \to \gamma\gamma/\gamma Z$ are already in tension with Fermi-LAT and HESS searches for line-like spectral features in the photon spectrum.

\subsubsection{Constraints from Fermi-LAT photon line searches}

Very recently, the Fermi-LAT collaboration has reported a constraint on a DM annihilation cross section based on updated searches for spectral line signatures in the energy range 200 MeV -- 500 GeV from around the Galactic Center (GC) using 5.8 years of data reprocessed with the Pass 8 event-level analysis~\cite{TheFermi-LAT:2015gja}.
In the analysis, they searched spectral lines expected from dark matter annihilation for four signal regions of interest (ROIs), selected to optimize sensitivity to different dark matter halo profiles: NFW with $\gamma=1$ or 1.3, Einasto, and isothermal.
In Ref.~\cite{TheFermi-LAT:2015gja}, it has been shown that the bound for $m_{\rm DM} \gtrsim 100$ GeV is just 2 -- 4 times weaker even for the isothermal profile
compared to the bound for the Einasto profile since the Fermi-LAT has measured gamma-rays from all the sky and thus can find the corresponding optimized ROI for each DM profile.
In this study, we use the $2\sigma$ upper limit on $\sigma v^{\gamma\gamma}$ for the Einasto profile with the local DM density $\rho_0=0.4$ GeV/cm$^3$ as the bound on the total cross section $\sigma v^{\gamma\gamma}$ + $\frac{1}{2} \sigma v^{\gamma Z} \equiv \sigma v_{\chi^0\chi^0 \to \gamma\gamma} + \frac{1}{2}\sigma v_{\chi^0\chi^0 \to \gamma Z}$, weighted by the number of photons for each final state.
In Figures~\ref{Fig-Higgsino}--\ref{Fig-Triplet1-BigGap}, we plot the constraint for the Einasto profile on $\sigma v^{\gamma\gamma}$ + $\frac{1}{2} \sigma v^{\gamma Z}$ as a blue-thick dotted line.

\subsubsection{Constraints from HESS photon line searches}

\begin{figure}[h]
	\begin{center}
		\includegraphics[width=13.2cm,height=8.5cm]{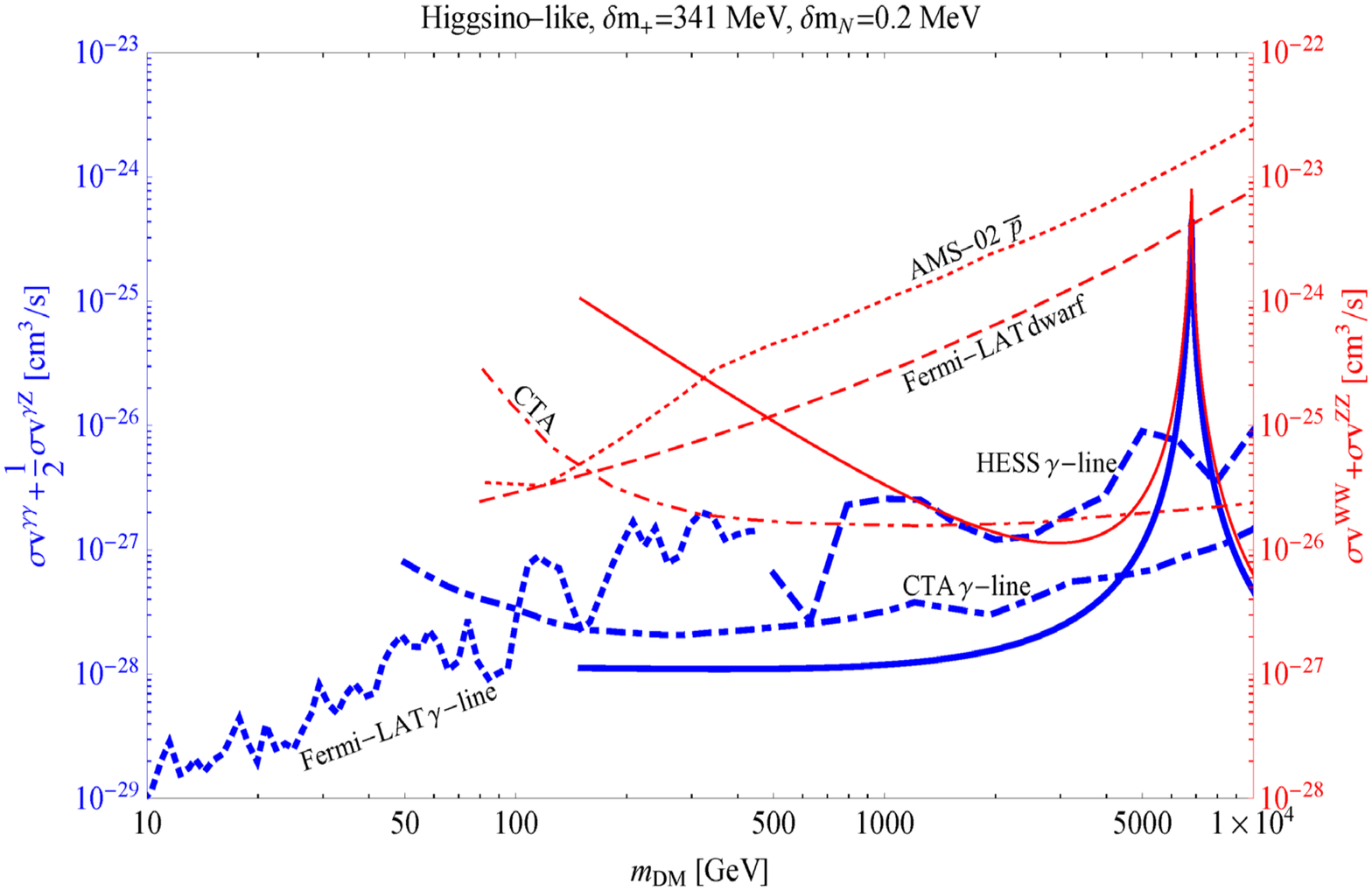}
        \includegraphics[width=13.2cm,height=8.5cm]{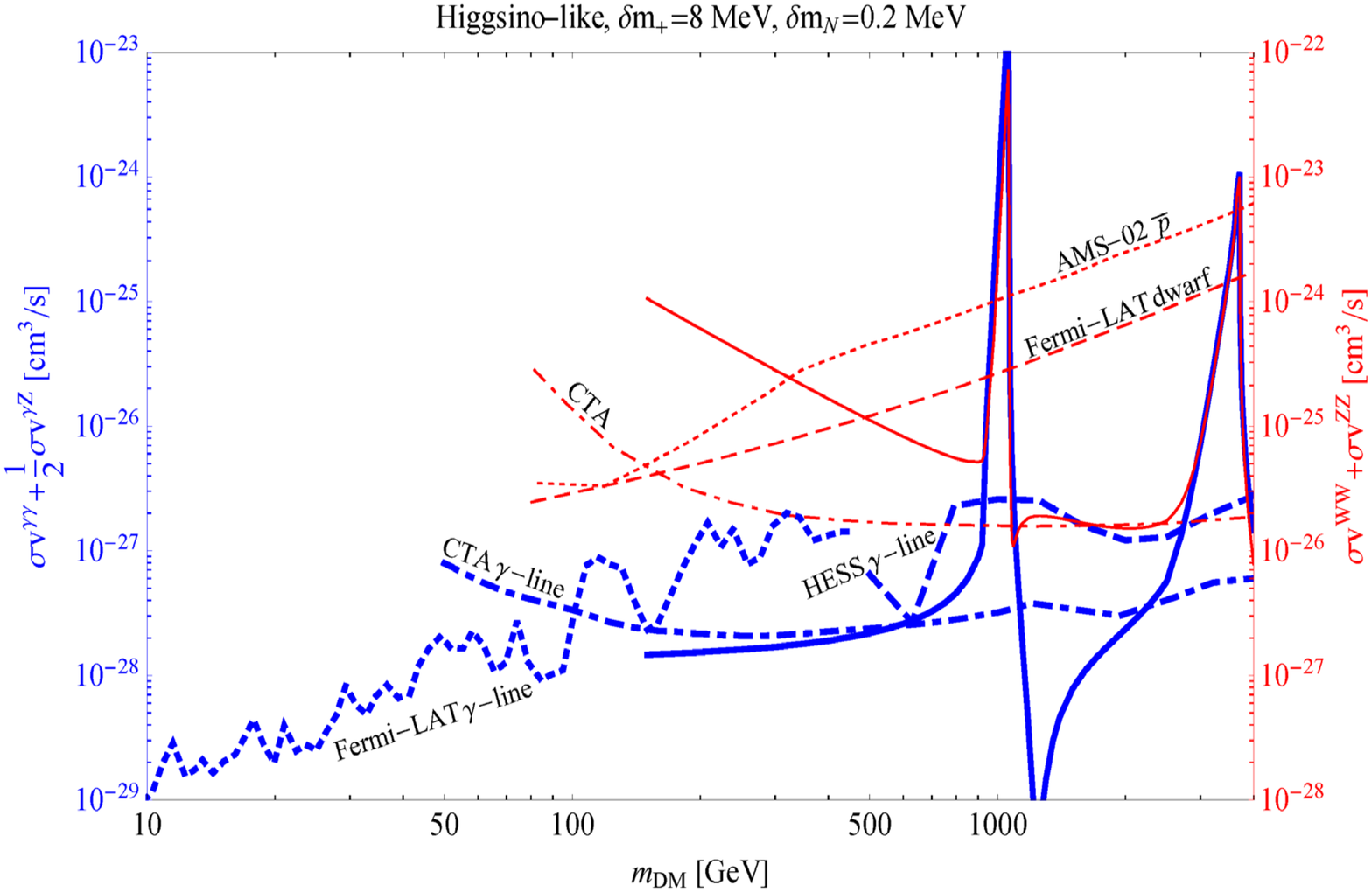}
	\end{center}
	\vspace*{-1.2cm}
\caption{Constraints on $\sigma v_{\chi^0\chi^0 \to W^+W^-} + \sigma v_{\chi^0\chi^0 \to ZZ}$ (red-thin curves with the scale on the right) and
$\sigma v_{\chi^0\chi^0 \to \gamma\gamma} + \frac{1}{2}\sigma v_{\chi^0\chi^0 \to \gamma Z}$ (blue-thick curves with the scale on the left) for the Higgsino-like EWDM with $\delta m_+=341$ MeV (top) and 8 MeV (bottom) for $\delta m_N=0.2$ MeV.
In each panel, the solid line is the calculated DM annihilation cross section including the non-perturbative effect.
The red-thin dotted and dashed lines show the upper limits obtained from the AMS-02 anti-proton flux data analysis~\cite{Giesen:2015ufa} and Fermi-LAT gamma-ray measurements from MW dwarf spheroidal satellite galaxies~\cite{Ackermann:2015zua}.
The blue-thick dotted and dashed lines are constraints from line-like photon signature searches by
Fermi-LAT~\cite{TheFermi-LAT:2015gja} and HESS~\cite{Abramowski:2013ax}, respectively.
The CTA sensitivities on the $WW/ZZ$~\cite{Lefranc:2015pza} and $\gamma X$~\cite{Bergstrom:2012vd, Ibarra:2015tya} channels are presented as red-thin and blue-thick dot-dashed curves, respectively.
}
\label{Fig-Higgsino}
\end{figure}

\begin{figure}[h]
	\begin{center}
        \includegraphics[width=13.2cm,height=8.5cm]{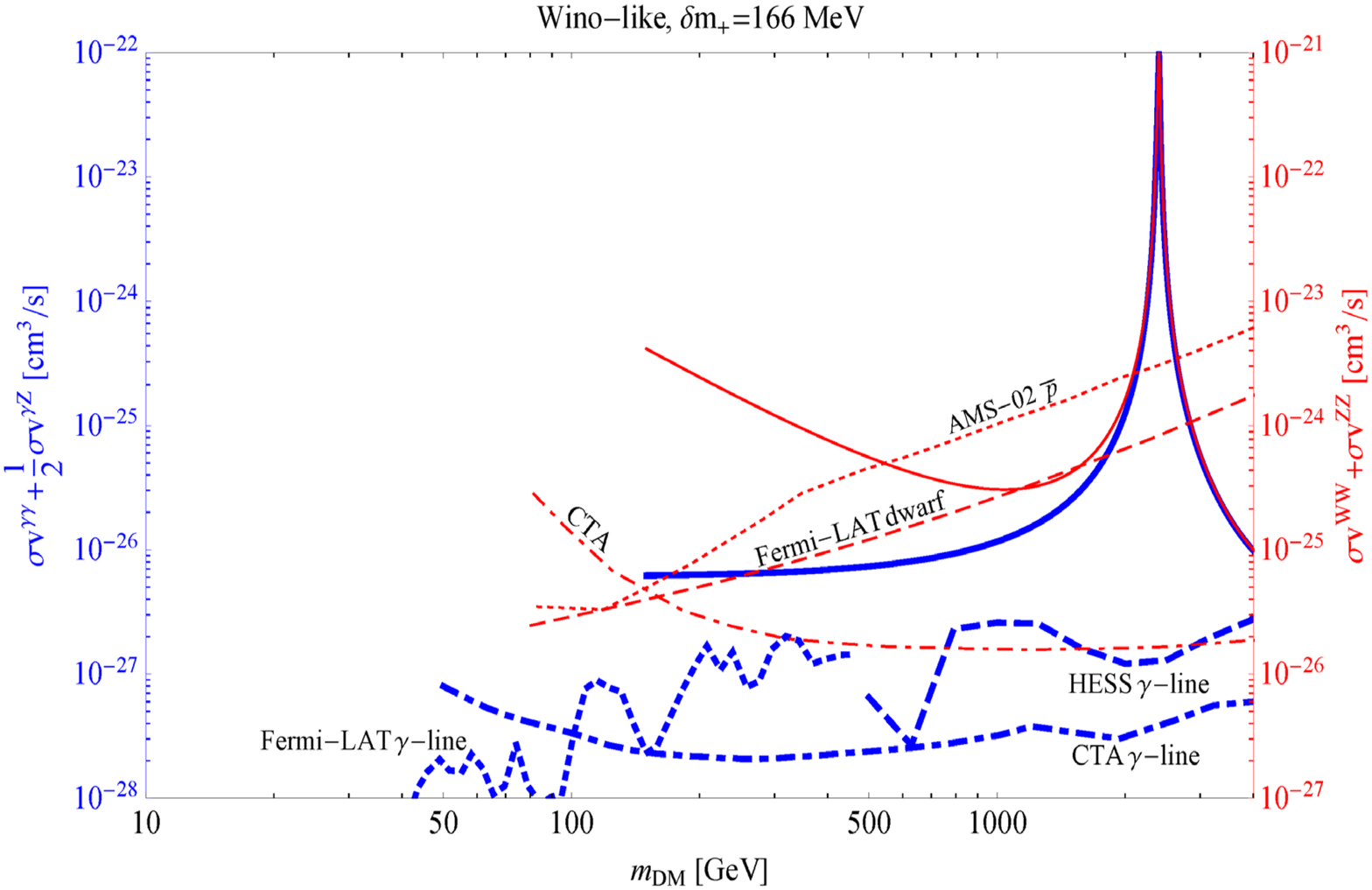}
        \includegraphics[width=13.2cm,height=8.5cm]{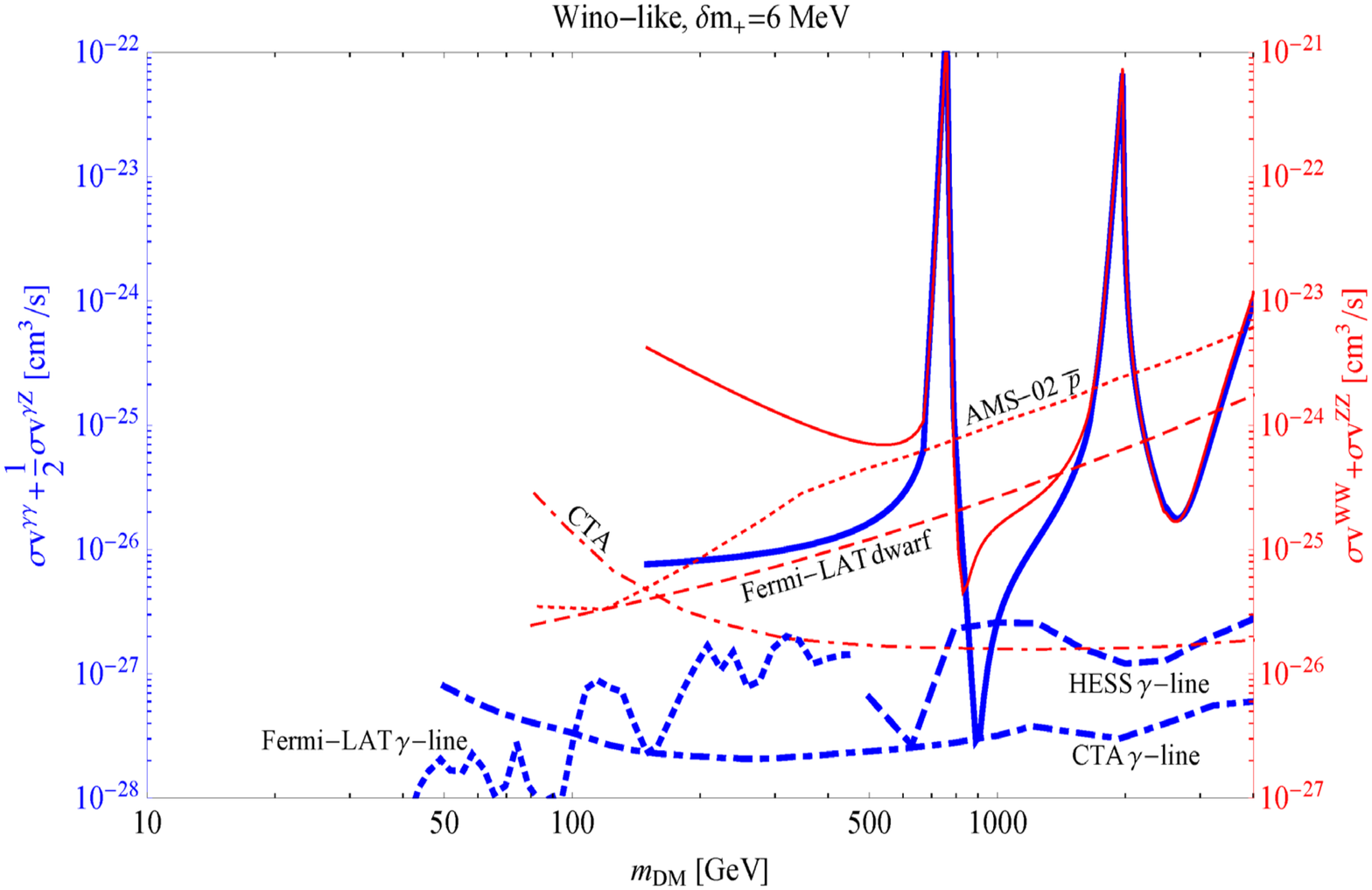}
	\end{center}
	\vspace*{-1.0cm}
\caption{Constraints  for the wino-like EWDM with $\delta m_+=166$ MeV (top) and 6 MeV (bottom).
Each line is the same as Figure\ref{Fig-Higgsino}.
}
\label{Fig-Wino}
\end{figure}

\begin{figure}[h]
	\begin{center}
		\includegraphics[width=13.2cm,height=8.5cm]{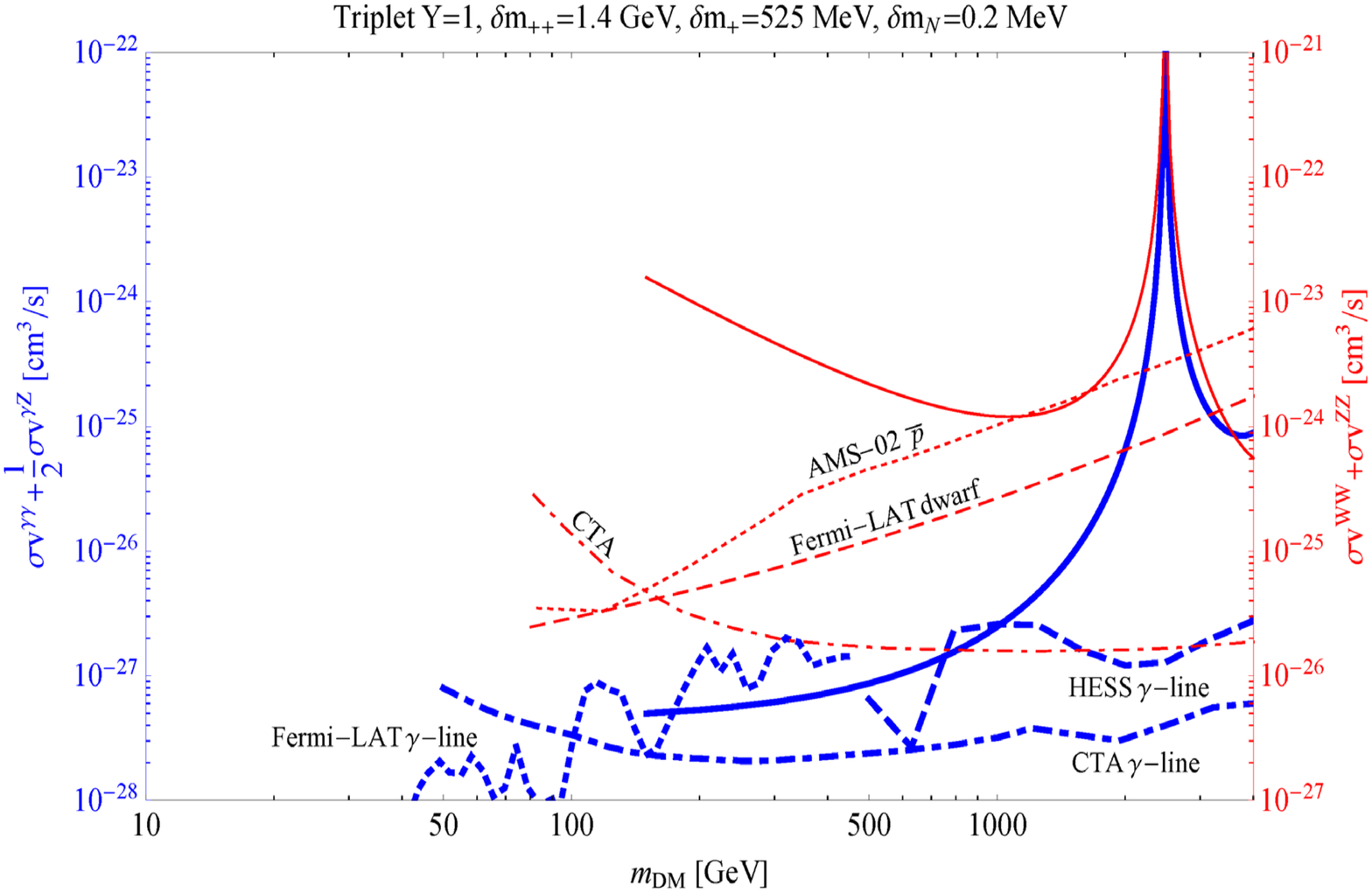}
        \includegraphics[width=13.2cm,height=8.5cm]{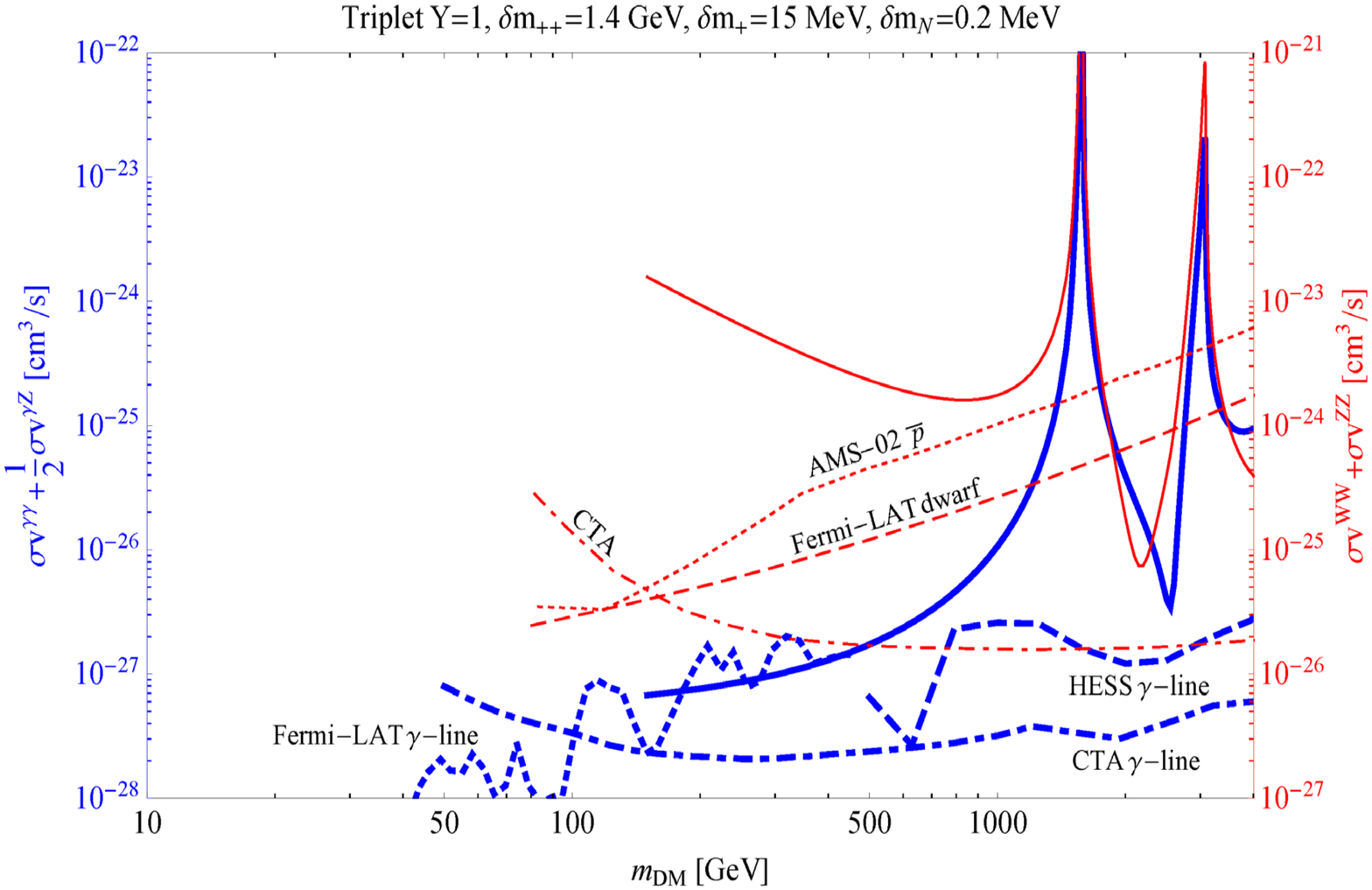}
	\end{center}
	\vspace*{-1.0cm}
\caption{Constraints for the triplet EWDM with $Y=\pm1$. $\delta m_+=525$ MeV (top-panel) and 15 MeV (bottom-panel) for $\delta m_{++}=1.4$ GeV and $\delta m_N=0.2$ MeV.
Each line is the same as Figure\ref{Fig-Higgsino}.
}
\label{Fig-Triplet1}
\end{figure}

\begin{figure}[h]
	\begin{center}
		\includegraphics[width=13.2cm,height=8.5cm]{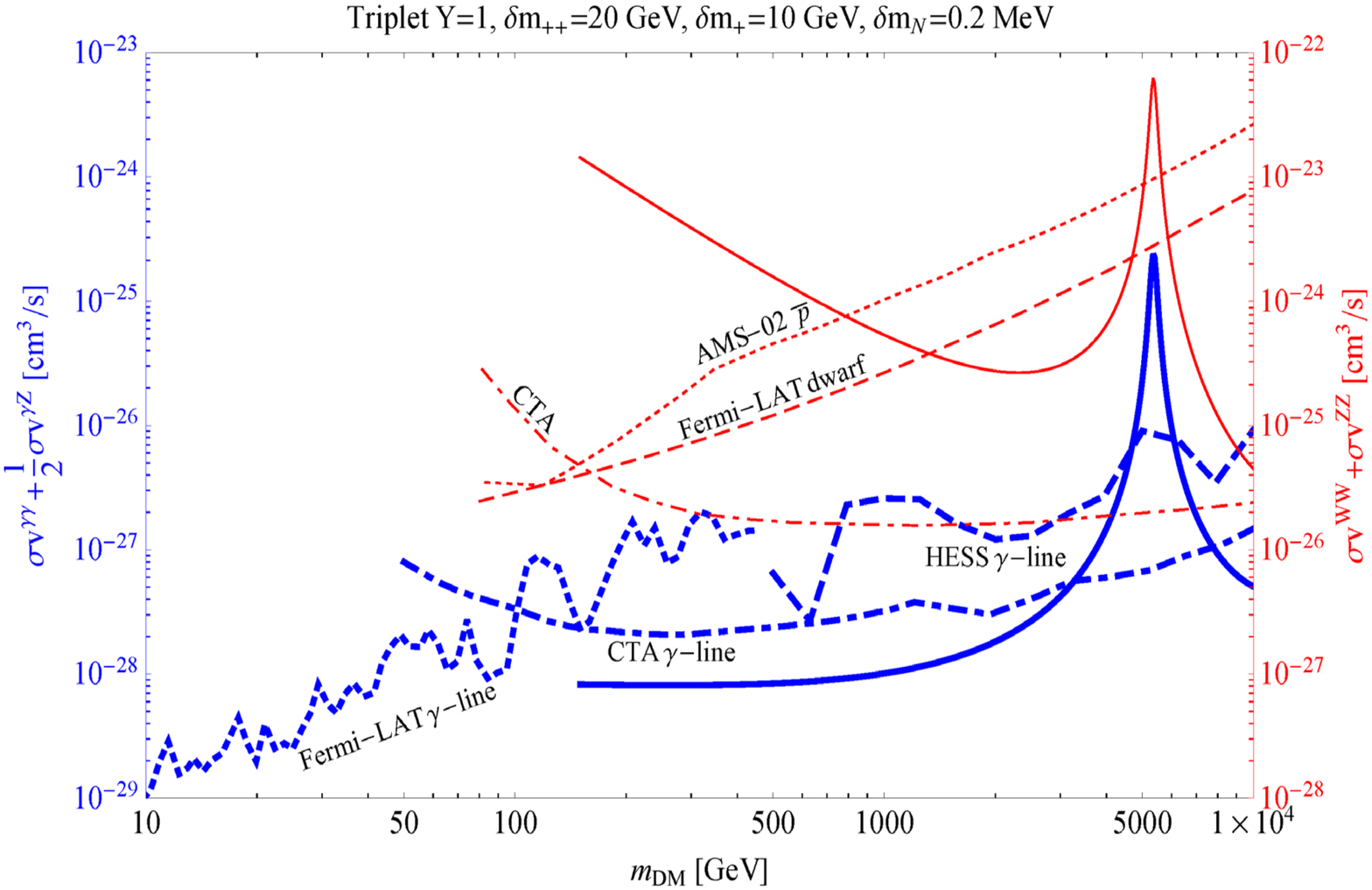}
	\end{center}
	\vspace*{-1.0cm}
\caption{Constraints for the triplet EWDM with $Y=\pm1$ for $\delta m_{++}=20$ GeV, $\delta m_+=10$ GeV, and $\delta m_N=0.2$ MeV.
Each line is the same as Figure\ref{Fig-Higgsino}.
}
\label{Fig-Triplet1-BigGap}
\end{figure}

In Ref.~\cite{Abramowski:2013ax}, upper limits on line-like gamma-ray signatures in the energy range 500 GeV -- 25 TeV are provided using the data collected by the
HESS, which complement the limits obtained by the Fermi-LAT at lower energies~\cite{TheFermi-LAT:2015gja}.
In the analysis, the HESS collaboration assumed the Einasto DM halo profile with $\rho_0=0.4$ GeV/cm$^3$.
However, the HESS collaboration has searched gamma-ray lines for only one ROI, a $1^\circ$ radius circle around the GC, compared to the Fermi-LAT.
Thus, the HESS limit for the Einasto profile can be weakened by about two orders of magnitude for a more cored profile, the isothermal profile.
In our analysis, we use the $2\sigma$ limit from the HESS spectral line search for the Einasto profile as a representative constraint on $\sigma v^{\gamma\gamma}$ + $\frac{1}{2} \sigma v^{\gamma Z}$ which is depicted as a blue-thick dashed line in Figures~\ref{Fig-Higgsino}--\ref{Fig-Triplet1-BigGap}.

%
%
%
%

%
%

\subsection{Summary of indirect constraints on EWDM}

We plot DM annihilation cross sections $\sigma v^{\gamma\gamma}$ + $\frac{1}{2} \sigma v^{\gamma Z}$ and $\sigma v^{WW} + \sigma v^{ZZ}$ as blue-thick and red-thin solid lines for the doublet, Majorana triplet, and hypercharged triplet ($Y=\pm1$) EWDMs in Figures~\ref{Fig-Higgsino}, \ref{Fig-Wino}, and \ref{Fig-Triplet1}, respectively.
For each type of EWDM, we calculate $\sigma v^{\gamma\gamma}$ + $\frac{1}{2} \sigma v^{\gamma Z}$ and $\sigma v^{WW} + \sigma v^{ZZ}$ as a function of DM mass $m_{\rm DM}$ including the non-perturbative effect for two representative values of $\delta m_+$: the typical mass splitting of $\mathcal{O}(0.1)$ GeV arising from the electroweak one-loop correction (top-panel) and $\mathcal{O}(10)$ MeV (bottom-panel).
For larger mass gap, the SRT effect becomes much weaker for a given DM mass.
In order to see this effect, we additionally show the results for the $Y=\pm1$ triplet EWDM with the mass gap of 10 GeV in Figure~\ref{Fig-Triplet1-BigGap}.
All the relevant indirect constraints on $\sigma v^{\gamma\gamma}$ + $\frac{1}{2} \sigma v^{\gamma Z}$ and $\sigma v^{WW} + \sigma v^{ZZ}$ are shown by blue-thick and red-thin curves, respectively.

\begin{itemize}
\item{\bf The limit from AMS-02 antiproton flux measurements}: We plot the $2\sigma$ limit on $\sigma v^{WW} + \sigma v^{ZZ}$ from AMS-02 antiproton flux measurements as a red-thin dotted line.
\item{\bf The limit from Fermi-LAT continuum photon searches}: For $\sigma v^{WW} + \sigma v^{ZZ}$, the upper region of the red-thin dashed curve is excluded by the limit from the Fermi-LAT continuum gamma-ray searches for the Milky Way satellite dwarf galaxies at the $2\sigma$ level.
\item{\bf The limit from Fermi-LAT photon line searches}: For $\sigma v^{\gamma\gamma}$ + $\frac{1}{2} \sigma v^{\gamma Z}$, the $2 \sigma$ exclusion limit from Fermi-LAT spectral line searches for around the GC is shown as a blue-thick dotted curve.
\item{\bf The limit from HESS photon line searches}: The HESS gamma-ray line signature search limit on $\sigma v^{\gamma\gamma}$ + $\frac{1}{2} \sigma v^{\gamma Z}$ is plotted as a blue-thick dashed curve.
\item{\bf The future sensitivity of CTA}: In near future, the remaining parameter regions for each EWDM will be probed by various upcoming cosmic-ray observations such as CTA~\cite{Acharya:2013sxa} and GAMMA-400~\cite{Galper:2014pua}.
    The CTA sensitivity with a 500 h time exposure on $\sigma v^{WW} + \sigma v^{ZZ}$~\cite{Lefranc:2015pza} is plotted as a red-thin dot-dashed line.
    For 5 h of GC observation with CTA, the upper limit on $\sigma v^{\gamma\gamma}$ + $\frac{1}{2} \sigma v^{\gamma Z}$~\cite{Bergstrom:2012vd, Ibarra:2015tya} is depicted as a blue-thick dot-dashed curve.
\end{itemize}

As stated in Introduction, each EWDM is assumed to account for 100 \% of the observed DM relic abundance in wide ranges of the DM mass and mass gaps.
In the case of the doublet (Higgsino-like) EWDM with a radiative mass splitting $\delta m_+ =341$ MeV, the DM mass larger than about 500 GeV is allowed except a narrow Sommerfeld peak region at 7 TeV.
For a smaller mass gap $\delta m_+=8$ MeV, the first Sommerfeld peak moves down to about 1 TeV and more peaks appear at lower DM masses.
These peak regions are excluded by either the Fermi-LAT gamma-ray data from the dwarf galaxies or HESS gamma-line data.
The Majorana triplet (wino-like) EWDM is stringently constrained for the whole range of masses up to $\sim 10$ TeV for the typical mass splitting of 166 MeV by the AMS-02, Fermi-LAT, and HESS results.
However, the indirect constraints on the wino-like EWDM could be evaded around Ramsauer-Townsend dips which can occur for a very small mass splitting,  $\mathcal{O}(10)$ MeV or less.
For the case of the triplet EWDM with $Y=\pm1$,\footnote{It is advocated as the unique candidate for asymmetric EWDM in Ref.~\cite{Boucenna:2015haa}, which is however stringently constrained by DM indirect detection limits as shown in this study.}
the indirect constraints become much more stringent to rule out for masses up to $\sim 20$ TeV by a combination of AMS-02, Fermi-LAT, and HESS limits.
It is also difficult to dodge the indirect constraints by arranging Ramsauer-Townsend dips with smaller mass splittings
as, contrary to the Sommerfeld peaks, the Ramsauer-Townsend dips for $\sigma v^{WW} + \sigma v^{ZZ}$ and $\sigma v^{\gamma\gamma}$ + $\frac{1}{2} \sigma v^{\gamma Z}$ do not coincide with each other as shown in the lower panel of Figure 3.
This behavior is more apparent in the case of the quintuplet EWDM
which exhibits non-overlapping narrow dips even with the radiative mass splitting \cite{Cirelli:2007xd}, and thus is completely ruled out by the combination of all the indirect constraints.
As shown in Figure~\ref{Fig-Triplet1-BigGap} taking a large mass splitting of $\mathcal{O}(10)$ GeV, the triplet EWDM with $Y=\pm1$ can be safe from all the current indirect constraints in the mass range of $m_{\rm DM} \gtrsim$ a couple of TeV except around the Sommerfeld peaks.
Similarly, the Wino-like EWDM can escape from the current indirect limits for $m_{\rm DM} \gtrsim 1$ TeV.
However, CTA~\cite{Acharya:2013sxa} will be able to probe most of remaining parameter regions even for the Higgsino-like EWDM and/or a large mass splitting of $\mathcal{O}(10)$ GeV.

\section{Conclusion}

In this paper, we have investigated indirect constraints on EWDM considering the SRT effect for its annihilations into a pair of SM gauge bosons which are sensitive to the size of mass splitting among the multiplet components.
Assuming that EWDM accounts for the observed DM relic density for a given DM mass and the radiative (or smaller) mass splitting, we found that the triplet with $Y=\pm1$ (and higher multiplet) EWDM is ruled out up to $m_{\rm DM} \approx $ 10 -- 20 TeV by the combination of current limits from AMS-02 (antiproton), Fermi-LAT (gamma-ray), and HESS (gamma-ray line) measurements, disregarding a potentially strong DM halo profile dependence of the HESS limit.
In the case of the Majorana triplet (wino-like) EWDM, there is a chance to dodge the indirect constraints around Ramsauer-Townsend dips only with a tiny mass splitting $\delta m _+ \lesssim \mathcal{O}(10)$ MeV.
On the other hand, the Higgsino-like EWDM is excluded just for DM masses less than $\sim 500$ GeV and around Sommerfeld resonance peaks.
Such a stringent limit can be weakened significantly if a large mass splitting of $\mathcal{O}(10)$ GeV is employed.
The indirect constraints could be evaded even for the wino-like and the $Y=\pm1$ triplet EWDM in the mass range of $m_{\rm DM} \gtrsim \mathcal{O}$(TeV) except around the Sommerfeld peaks which occur at larger DM masses.
However, the unconstrained parameter regions will be mostly searched by various future cosmic-ray measurements such as CTA~\cite{Acharya:2013sxa} and GAMMA-400~\cite{Galper:2014pua}.

\vspace{0.5 cm}
\begin{acknowledgements}
JCP is supported by Basic Science Research Program through the National Research Foundation of Korea funded  by the Ministry of Education (NRF-2013R1A1A2061561),
and appreciate CETUP* (Center for Theoretical Underground Physics and Related Areas) for its hospitality during completion of this work.
We thank Shigeki Matsumoto for useful discussions on the HESS bound.
\end{acknowledgements}

\begin{appendix}
\section{The potential and tree-level annihilation matrices for EWDMs}\label{appendix}

Considering the covariant derivative $D_\mu =
\partial_\mu + i g_2 A_\mu T^A$ for each gauge boson $A=W^\pm, Z, \gamma$, the
potential matrix in Eq.~(\ref{schroedinger}) and the tree-level
scattering matrix $\Gamma_{ij}$ take the following general forms:
\begin{eqnarray}
&& V_{ij}(r) = 2 \,\delta m_{i}\, \delta_{ij} - \alpha_2 N_i N_j
 \sum_{A} \left[ T^A_{ij}\right]^2 {e^{-m_A r} \over r}
\quad \mbox{with} \quad \delta m_{i} = m_{\chi_i} - m_{\chi_0}\,; \\
&&  \Gamma_{ij}^{AB} = {\pi \alpha_2^2 \over 2(1+\delta_{AB}) m^2_{\rm DM}} f(x_A,x_B)
 N_i N_j \left\{ T^A, T^B \right\}_{ii}
  \left\{ T^A, T^B \right\}_{jj}\,, \\
&& \mbox{where}~~ f(x_A,x_B) \equiv { \left(1-{x_A+x_B \over 2}\right) \over
 \left( 1 - {x_A+x_B \over 4} \right)^2 }
 \sqrt{ 1 - {x_A+x_B \over 2} +{( x_A - x_B)^2 \over 16}} ~~\mbox{with}~~
 x_A = {m_A^2 \over m^2_{\rm DM} }\,.\nonumber
\end{eqnarray}
Here the normalization factor $N_i$ is  $1$ or $ \sqrt{2}$ for the
Dirac (charged) or Majorana (neutral) two-body state,
respectively. In the following, we present the explicit matrix elements for the three fermionic EWDM candidates in the lowest multiplets.

\begin{itemize}
\item Doublet (Higgsino-like) EWDM

The potential matrix $V$ and the normalized tree-level scattering matrix $\Gamma_{AB}$
normalized by  $(\pi \alpha_2^2 / m_{\rm DM}^2) f(x_A,x_B)$
in the basis of the three states $( \chi^+ \chi^-, \chi^0_1 \chi^0_1, \chi^0_2 \chi^0_2)$ are given as follows:
\begin{eqnarray}
V &=&
\begin{pmatrix}
\mbox{\small $2\, \delta m_{+}$} -{s_W^2 \alpha_2  \over r}
-{(1-2 s_W^2)^2 \alpha_2 \over 4 c_W^2 } { e^{-m_Z r} \over r} &
- {\alpha_2 \over 2\sqrt{2}} { e^{-m_W r} \over r} &
- {\alpha_2 \over 2\sqrt{2} } {e^{-m_W r} \over r}  \\
- {\alpha_2 \over 2\sqrt{2} } {e^{-m_W r} \over r} &
0 &
-{\alpha_2 \over 4 c_W^2 } {e^{-m_Z r} \over r} \\
- {\alpha_2 \over 2\sqrt{2}} {e^{-m_W r} \over r} &
-{\alpha_2 \over 4 c_W^2 } {e^{-m_Z r} \over r}&
\mbox{\small $2\, \delta m_{N} $}
\end{pmatrix} ;
\end{eqnarray}
\begin{eqnarray}
\Gamma_{WW} & =& {1\over 16}
\begin{pmatrix}
2 &  \sqrt{2} &  \sqrt{2} \\
 \sqrt{2} & 1 & 1 \\
 \sqrt{2} & 1 & 1
\end{pmatrix} ,
 \\
\Gamma_{ZZ} & =& {1\over 128 c_W^4}
\begin{pmatrix}
8 (1-2 s_W^2)^4 &  2 \sqrt{2} (1-2 s_W^2)^2 &  2 \sqrt{2} (1-2 s_W^2)^2 \\
 2 \sqrt{2} (1-2 s_W^2)^2  & 1 & 1 \\
 2 \sqrt{2} (1-2 s_W^2)^2   & 1 & 1
\end{pmatrix} ,
\nonumber\\
\Gamma_{\gamma Z} & =& {s_W^2 (1-2 s_W^2)^2 \over  2 c_W^2}
\begin{pmatrix}
1 & 0 & 0 \\
0 & 0 & 0 \\
0 & 0 & 0
\end{pmatrix} , \quad
\Gamma_{\gamma \gamma}  = {s_W^4 }
\begin{pmatrix}
1 & 0 & 0 \\
0 & 0 & 0 \\
0 & 0 & 0
\end{pmatrix} .
\nonumber
\end{eqnarray}

\item Majorana triplet (Wino-like) EWDM

The potential matrix $V$ and the tree-level scattering matrix $\Gamma_{AB}$
normalized by  $(\pi \alpha_2^2 / m_{\rm DM}^2) f(x_A, x_B)$
in the basis of the two states $( \chi^+ \chi^-, \chi^0 \chi^0)$ are given as follows:
\begin{eqnarray}
V &=&
\begin{pmatrix}
\mbox{\small $2\, \delta m_{+}$}
-{s_W^2 \alpha_2  \over r}  - {{c_W^2 \alpha_2 } e^{-m_Z r} \over r} &
-  {{ \sqrt{2}\alpha_2} e^{-m_W r} \over r} \\
-  { { \sqrt{2}\alpha_2}e^{-m_W r} \over r} & 0
\end{pmatrix} ;
\end{eqnarray}
\begin{eqnarray}
\Gamma_{WW} & =& {1\over 2}
\begin{pmatrix}
1 &  \sqrt{2} \\
 \sqrt{2} & 2
\end{pmatrix},
\quad
\Gamma_{ZZ}  =
\begin{pmatrix}
c_W^4 & 0 \\
0 & 0
\end{pmatrix} ,
\\
\Gamma_{\gamma Z} & =&
\begin{pmatrix}
2 s_W^2 c_W^2 & 0 \\
0 & 0
\end{pmatrix} , \quad
\Gamma_{\gamma \gamma}  =
\begin{pmatrix}
{s_W^4 }  & 0 \\
0 & 0
\end{pmatrix} .
\nonumber
\end{eqnarray}
Let us note that there are a few discrepancies in factors of the scattering matrices (A4) and (A6)
compared with the previous results in Ref.~\cite{Hisano:2004ds}.

\item Hypercharged triplet EWDM

The potential matrix $V$ and the tree-level scattering matrix $\Gamma_{AB}$
normalized by  $(\pi \alpha_2^2 / m_{\rm DM}^2) f(x_A,x_B)$
in the basis of the four states $(\chi^{++}\chi^{--}, \chi^+ \chi^-, \chi^0_1 \chi^0_1, \chi^0_2 \chi^0_2)$ are given as follows:
\begin{equation}
V=
\begin{pmatrix}
\mbox{\small $2\, \delta m_{++}$} -{4 s_W^2 \alpha_2  \over r}
-{(1-2 s_W^2)^2 \alpha_2 \over c_W^2 } { e^{-m_Z r} \over r} &
-{ \alpha_2  e^{-m_W r} \over r} & 0 & 0 \\
-{ \alpha_2  e^{-m_W r} \over r} &
\mbox{\small $2\, \delta m_{+}$}  -{ s_W^2 \alpha_2  \over r}
-{s_W^4 \alpha_2 \over c_W^2 } { e^{-m_Z r} \over r} &
-{\alpha_2 \over \sqrt{2}}  { e^{-m_W r} \over r} &
-{\alpha_2 \over \sqrt{2}}  { e^{-m_W r} \over r} \\
0 & {\alpha_2 \over \sqrt{2}}  { e^{-m_Z r} \over r} &
0 & {\alpha_2 \over c_W^2}  { e^{-m_Z r} \over r} \\
0 &
- {\alpha_2 \over \sqrt{2} } {e^{-m_Z r} \over r}  &
- {\alpha_2 \over c_W^2 } {e^{-m_Z r} \over r} &
\mbox{\small $2\, \delta m_{N} $}
\end{pmatrix} ;
\end{equation}
\begin{eqnarray}
&&\Gamma_{WW}={1\over 4}
\begin{pmatrix}
2 & 4 & \sqrt{2} &  \sqrt{2} \\
4 & 8 & 2\sqrt{2} & 2 \sqrt{2} \\
\sqrt{2} & 2 \sqrt{2} & 1 & 1 \\
 \sqrt{2} & 2 \sqrt{2} & 1 & 1
\end{pmatrix} ,
\\
&&\Gamma_{ZZ}={1\over 8 c_W^4}
\begin{pmatrix}
2 (1-2 s_W^2)^4 &  8  (1-2 s_W^2)^2 s_W^4 &  2 \sqrt{2} (1-2 s_W^2)^2 & 2 \sqrt{2} (1-2 s_W^2)^2 \\
 8 (1-2 s_W^2)^2 s_W^4 & 8 s_W^8 & 2 \sqrt{2} s_W^4 &  2 \sqrt{2} s_W^4 \\
  2 \sqrt{2} (1-2 s_W^2)^2 & 2 \sqrt{2} s_W^4 & 1 & 1 \\
 2 \sqrt{2} (1-2 s_W^2)^2 &   2 \sqrt{2} s_W^4 & 1 & 1
\end{pmatrix} ,
\nonumber\\
&&\Gamma_{\gamma Z}=2 { s_W^2 \over c_W^2}
\begin{pmatrix}
4 (1-2 s_W^2)^2 & -4 (1-2 s_W^2)^2 s_W^2 & 0 & 0 \\
-4 (1-2 s_W^2)^2 s_W^2 & s_W^4 & 0 & 0 \\
0 & 0 & 0 & 0 \\
0 & 0 & 0 & 0
\end{pmatrix} , \quad
\Gamma_{\gamma \gamma}  = {s_W^4 }
\begin{pmatrix}
16 & 4 & 0 & 0 \\
4   & 1 & 0 & 0 \\
0 & 0 & 0 & 0 \\
0 & 0 & 0 & 0
\end{pmatrix} .
\nonumber
\end{eqnarray}

\end{itemize}
\end{appendix}

\end{document}